\documentclass{bmcart}

\usepackage[utf8]{inputenc} 

\usepackage{graphicx}
\usepackage{subfigure}
\usepackage{multicol}
\usepackage{amsmath}

\usepackage{tikz,xcolor,hyperref}
\newcommand{\revision}[1]{\textcolor{blue}{#1}}

\startlocaldefs
\endlocaldefs

\begin{document}

\begin{frontmatter}

\begin{fmbox}
\dochead{Research}


\title{A Combined Synchronization Index for Grassroots Activism on Social Media}


\author[
   addressref={aff1},                   
   corref={aff1},                       
   email={lynnetteng@cmu.edu}   
]{\inits{LHXN}\fnm{Lynnette Hui Xian} \snm{Ng}}
\author[
   addressref={aff1},
   email={carley@andrew.cmu.edu}
]{\inits{KMC}\fnm{Kathleen M} \snm{Carley}}


\address[id=aff1]{
  \orgname{CASOS, Software and Societal Systems}, 
  \street{4665 Forbes Ave},                     %
  \postcode{15213}                                
  \city{Pittsburgh},                              
  \cny{USA}                                    
}


\begin{artnotes}
\end{artnotes}

\end{fmbox}


\begin{abstractbox}

\begin{abstract} 
Social media has provided a citizen voice, giving rise to grassroots collective action, where users deploy a concerted effort to disseminate online narratives and even carry out offline protests. Sometimes these collective action are aided by inorganic synchronization, which arise from bot actors. It is thus important to identify the synchronicity of emerging discourse on social media and the indications of organic/inorganic activity within the conversations. This provides a way of profiling an event for possibility of offline protests and violence.
In this study, we build on past definitions of synchronous activity on social media -- simultaneous user action -- and develop a Combined Synchronization Index (CSI) which adopts a hierarchical approach in measuring user synchronicity. We apply this index on six political and social activism events on Twitter and analyzed three action types: synchronicity by hashtag, URL and @mentions.The CSI provides an overall quantification of synchronization across all action types within an event, which allows ranking of a spectrum of synchronicity across the six events. Human users have higher synchronous scores than bot users in most events; and bots and humans exhibits the most synchronized activities across all events as compared to other pairs (i.e., bot-bot and human-human). We further rely on the harmony and dissonance of CSI-Network scores with network centrality metrics to observe the presence of organic/inorganic synchronization. We hope this work aids in investigating synchronized action within social media in a collective manner.




\end{abstract}


\end{abstractbox}
%

\end{frontmatter}


\section*{Introduction}
Social media has facilitated mainstream audiences to have a voice in political and social issues. The construction of citizen voice in these discussions has been established as civic values \cite{dumitrica2020voice}. However, the multiplicity of voices and opinions on online social networks can exhibit characteristics of complex synchronization stemming from grassroots collective action. These grassroots efforts deploy concerted efforts to influence political decision making through dispersion of online narratives. When they elicit strong emotions, such campaigns can be a precursor of offline protests~\cite{jarvis2021my}. These protests, riots and other forms of violence threaten the safety of our social fabric, and thus early detection of the emergence and intensity of online synchronization is critical.

From the Arab Spring anti-government protests in 2011 to the Capitol Riots in 2021, social media has played a role in facilitating the organization and activism of grassroots action. These events are key topics that people have strong opinions about as they can result in a change in society, be it a change in government or existing laws \cite{goodwin2001passionate}. Thus, they are willing to invest resources to try and reshape their opportunity structure, sometimes inducing societal change in the process \cite{lazega2016synchronization}. We observe a portion of these synchronized online activity within the social media space, and seek to quantify it.

Synchronization on social media involves the identification of online actions by different actors that are synchronous in time. The study of synchronization on the web is particularly of interest for political and social activism events. 
Grassroots collective action refers to a specific type of synchronization, in which users disperse consistent messages in activist events \cite{lazega2016synchronization}. 
These collective information dissemination activities can segregate the online discourse into multiple, sometimes polarizing, narrative groups \cite{websci2022}. Polarization on social media creates and maintains echo chambers, which reduces the amount of diversity in the user experience \cite{pansanella2022modeling}. One important part of the conversation on coordinated action is how organic the action is. Inorganic activity stems from the presence of bot activity, which has been shown to manipulate the online discourse in social issues like democracy and activism \cite{strudwicke2020junkscience}. 


Bots have been observed to be active participants in political and social discussions. Their participation in grassroots conversations is of particular concern, because these automated accounts have been demonstrated to influence human users through content generation and social interactions, while successfully keeping their identity hidden \cite{messias2013you,ferrara2015manipulation}.

During the days after President Trump's announcement of the US withdrawal from the Paris Agreement, bots produced up to 25\% of original tweets on denialist discourses, questioning the importance of climate change \cite{marlow2021bots}. Social bots have also been studied in political discussions. In one study, they have been shown to distort the 2016 US Presidential elections, gathering political support and employing influence mechanisms such as creating information campaigns explicitly devoted to target the candidate of one’s opposing political leaning \cite{bessi2016social}.


Algorithms measuring synchronized activities on social media typically involve detecting users that perform high levels of the same, or similar, action within a specific time window. This temporal-based time windowing approach relies on inferred links between users by their observed user actions.
The links can be constructed based on similar retweets \cite{9381418}, similar URLs \cite{cao2015organic}, similar hashtags \cite{weber2021amplifying} or even similar texts \cite{ng2021coordinating}. After detection of synchronized users, the next step typically involves building a weighted user-similarity network \cite{pacheco2021uncovering}, before performing network clustering techniques ranging from Louvain community detection to focal structures analysis to extract the coordinating user communities \cite{websci2022,weber2021amplifying}. 
In using this technique to identify synchronized users via similarity of hashtags within the Hong Kong protest event in 2018, \cite{pacheco2021uncovering} discovered two key communities: an anti-protest group that shares images with Chinese texts and a pro-protest group that shares images with English text, showcasing the efficacy of this technique in identifying communities of polarizing intent.

However, these synchronization detection frameworks analyze users across only one action dimension at a time, rather than all action types in a combined form. For example, only the action of posting with a URL is analyzed in the Hong Kong protest case study, and the results are not combined with the coordination observations of other action types. In addition, the volume of filtered users returned can still be large: i.e., \cite{DBLP:journals/corr/abs-2105-07454} identifies 430k out of 628k users as coordinating with each other \cite{websci2022}, which is still a voluminous amount of users to investigate.

In the present work, we propose and develop a combined index for measuring synchronization across several dimensions of action types within datasets collected from Twitter. This index serves as a tool that fills an existing methodological gap, in which synchronization is analyzed by frequency of simultaneous actions and visual analysis. In comparison to the methods used in prior work, we show that a combined index of the synchronization networks provides the unique ability to span multiple dimensions of the analysis of online synchronization. This specification provides a high-level assessment of the synchronicity of users within an event through the network-based synchronization measure, then provides a deeper dive into the user and user-pair synchronization measures.

\subsection*{Contributions}
This paper builds on past work which developed techniques for measuring synchronization through high frequency of the same actions \cite{websci2022,9381418}. In these previous works, separate analyses of synchronicity via different actions were analyzed. This work then extends previous work on synchronization to provide an overall computation of the synchronization of a network, allowing quantitative ranking of network synchronizations. 
The contributions of this paper are the following: (1) We propose a hierarchical Combined Synchronization Index (CSI) to provide measures of user-pair, user and network synchronization of the discourse formed during an event. While there have been works that measure synchronization on social media, we are unaware of any works that provide a consolidated index to analyzing the synchronous activities of online users.
(2) We evaluate the nature of synchronization in terms of how organic the synchronization is. To do so, we identify inauthentic bot actors and compare the extent of synchronization between bot and human users. 
(3)~We quantitatively analyze the synchronization patterns and the nature of synchronization within six datasets relating to social activism and political events, drawing observations of their similarities and differences.

\section*{Combined Synchronization Index (CSI)}
We define a Combined Synchronization Index (CSI) to characterize the degree of synchronization between users that tweeted within an event. The discussion surrounding each event is a sub-component or subset derivation of the event's entire social network. We propose the Combined Synchronization Index, a hierarchical index, which begins with CSI-UserPair, then aggregating the values to obtain CSI-User for each user, and finally a singular value for CSI-Network.

\begin{figure}[h!]
\includegraphics[width=1.0\textwidth]{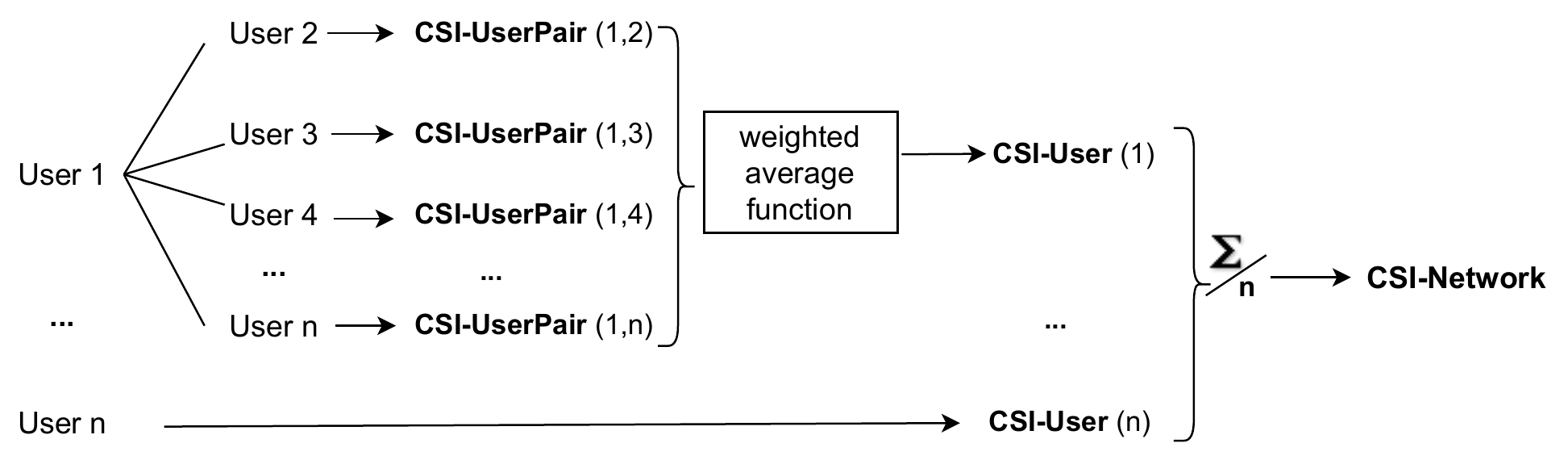}
\caption{\csentence{Computation of Combined Synchronization Index}
  Flowchart of the computation and aggregation of the hierarchical Combined Synchronized Index.}
\label{fig:csi_figure}
\end{figure}

CSI-UserPair quantifies the extent of synchronization between a pair of two users. For each user, the CSI-UserPair values for that user represents the other users he synchronizes with. CSI-User is an aggregated weighted average function, which puts higher weight on the values of user pairs that synchronize more frequently with each other. Finally, the CSI-User values are averaged out to obtain a CSI-Network value, which quantifies the extent of synchronicity for the slice of social network obtained during the event data collection. The concept of the Combined Synchronization Index is illustrated in Figure \ref{fig:csi_figure}. The CSI is measured in terms of number of times users synchronize, but for brevity we shall measure CSI in terms of \textit{points}.

\subsection*{CSI-UserPair}
CSI-UserPair provides an indication of how much two users $u$ and $v$ synchronize with each other, expressing the same semantic or social information in their tweets. Before calculating CSI-UserPair values, we first find users that synchronize with each other. In this work, we define synchronization between two users when they perform an action on social media at the similar point in time \cite{DBLP:journals/corr/abs-2105-07454,9381418}. An action is creating a Twitter post with one of the three post artifacts which we call "Action Types": URL, hashtag or @mention. In our study, we use all three action types, measuring synchronicity under different dimensions. We relax the temporal criteria and find synchronicity within a 5-minute time window, instead of a single point in time. The 5-minute interval is short enough to capture deliberate actions rather than coincidental synchronicity.

With this formulation, we obtain user pairs with a count of the number of times they synchronize with each other on each specific action, returning three sets of user pairs. This count, $S(u,v)$, is stored for the subsequent calculation of CSI-User. We normalize this count along the axis of each action, before collectively examined to generate CSI-UserPair data. The sets of user pairs are collectively examined to generate CSI-UserPair data. Where there are the same user pairs between these sets, the counts are summed together. To account for duplicate user pair counts across action types $a_i$, the number of action types $|a|$ that the user pair are found in synchrony is subtracted from the normalized summation. Then, to scale the user pair value according to the extent of synchronization, $|a|$ is multiplied to the final value. The calculation of CSI-UserPair is represented in Equation \ref{eq:csi-userpair}.

\begin{equation}
\begin{aligned}
   &  \textrm{CSI-UserPair}(u,v) = [\textrm{CSI-UserPair}(u,v,a_1) + \textrm{CSI-UserPair}(u,v,a_2) + \textrm{CSI-UserPair}(u,v,a_3)] \\ 
   & - |a| * |a|\\
   & \textrm{where } a_i \textrm{ is the action types users $u$, $v$ synchronize in}
\end{aligned}
    \label{eq:csi-userpair}
\end{equation}

Intuitively, the more times users perform a synchronized action, the higher they should rank on a synchronous index. Similarly, users should also rank higher if they synchronize in more action types $a$.

\subsection*{CSI-User}
The CSI-User value is a representation of the degree of synchronization a user partakes in within a network. It is a summation of CSI-UserPair values that a user $u$ is part of, weighted by the frequency of synchronicity of the paired user $v$. Intuitively, if users $u$ and $v$ synchronize with each other a lot, their CSI-UserPair(u,v) value should contribute a higher weight to the synchronicity measure of user $u$. CSI-User provides a quantitative way to  rank users according to their scale of synchronization with respect to other users in the network. The calculation of CSI-User is represented in Equation \ref{eq:csi-user}.

\begin{equation}
\begin{aligned}
   &  \textrm{CSI-User}(u) = \sum_{i=1}^n (S(u,v) * \textrm{CSI-UserPair(u,v)} ) \\
   & \textrm{for all $n$ pairs that user $u$ is part of}
\end{aligned}
    \label{eq:csi-user}
\end{equation}

\subsection*{CSI-Network}
A network can be characterized by CSI-Network, calculated by averaging the CSI-User values. This reveals the average amount of synchronization between users surrounding an event. The calculation of CSI-Network is revealed in Equation \ref{eq:csi-network}. 

\begin{equation}
\begin{aligned}
   &  \textrm{CSI-Network} = \sum^{n}_{i=1}(\textrm{CSI-User})/ n \\
   & \textrm{where } n \textrm{ is the number of synchronizing users in a network}
\end{aligned}
    \label{eq:csi-network}
\end{equation}

\section*{Data}
We measured user synchronicity across six Twitter datasets. We focused on social activism and political events, selecting for discourse that has a strong stance, in which synchronization may be used to champion the cause. With these parameters, we examine four social activism events: (1) Black Panther movie release in 2018; (2) Charlie Hebdo shooting anniversary in 2020; (3) ReOpen America protests in 2020; (4) COVID Vaccine Release in 2021. We also examine two political events: (1) US Election Primaries in 2020, and (2) Capitol Riots in 2021.

We used the Twitter V1 API for data collection, using the API's streaming endpoint to collect defined hashtags across a specified time frame. We filtered for Tweets in the English language to facilitate downstream analysis. To reduce noise, we analyze only tweets that are deliberately written by a Twitter user. To do so, we filtered the initial set of collected tweets to retain only the original tweets, i.e. tweets that are not retweets, quote tweets or replies. In total, the datasets had about 1.8 million unique users and 4.3 million original tweets. A summary of the data collection parameters and its corresponding statistics is reflected in Table \ref{tab:dataset}.


\begin{itemize}
    \item[] \underline{Social Activism Events}
    \item[] \textbf{Black Panther movie release 2018} follows responses that promoted Black power to Marvel Studio's Black Panther movie.
    \item[] \textbf{Charlie Hebdo shooting 2020} follows the discourse when suspects involved in the Charlie Hebdo shooting were tried in France. The shooting was a militant terrorist attack in Paris in 2015.
    \item[] \textbf{ReOpen America protests 2020} were launched across the United States against government-imposed lockdown measures in order to curb the global COVID-19 pandemic.
    \item[] \textbf{COVID Vaccine Release 2021} is a pivotal event in the global COVID-19 pandemic, as it meant that there is hope of a societal and economic recovery.
    \item[] \underline{Political Events}
    \item[] \textbf{US Election Primaries 2020} follows the discourse of the Primaries of the US 2020 Elections. This election ended with Joe Biden as the 46th President of the United States.
    \item[] \textbf{Capitol Riots 2021} happened on the 6th January 2021, where a mob stormed the US Capitol Building, claiming electoral fraud and that the Democrats stole the elections.
\end{itemize}

\begin{table}[!hpt]
\centering
\begin{tabular}{|p{4cm}|p{4cm}|p{3.0cm}|p{1.8cm}|p{1.8cm}|}
\hline
\textbf{Event} & \textbf{Collection Hashtag} & \textbf{Timeframe} & \textbf{Unique Users} & \textbf{Original Tweets} \\ \hline 
\multicolumn{5}{|c|}{Social Activism Events} \\ \hline
Black Panther 2018 & \#BlackPanther \#BlackHistoryMonth \#blackpower \#Afrofuturism & 8 Feb - 23 Feb 2018 & 75,752 & 105,910\\ \hline 
CharlieHebdo shooting 2020 & \#charliehebdo \#macron \#prophetmuhammad \#boycottfrance \#boycottfranceproducts \#macronapologizeToMuslims \#Islamophobia \#France\_spreadinghate & 3 Aug - 9 Nov 2020 & 219,188 & 643,956 \\ \hline
Reopen America 2020 & \#operationgridlock \#reopenamerica & 1 April - 22 June 2020 & 88,768 & 200,631\\ \hline
COVID Vaccine Release 2021 & \#vaccinesavelives \#vaccinekills \#takethevaccine \#notovaccine & 9 Jan - 9 May 2021 & 668,309 & 1,279,919\\ \hline
\multicolumn{5}{|c|}{Political Events} \\ \hline 
US Elections Primaries 2020 & \#uselectionsprimaries & 1 Feb -13 Feb 2020 & 628,705 & 2,223,056\\ \hline 
Capitol Riots 2021 & \#stopthesteal \#electionfraud \#makeamericagreatagain & 1 Jan - 7 Jan 2021 & 412,788& 763,850\\ \hline
\end{tabular}
\caption{Summary of data used in this study}
\label{tab:dataset}
\end{table}

\section*{Methodology}
For each of the six event networks, we first calculated the hierarchies of CSI -- CSI-UserPair, CSI-User and CSI-Network. We used the information to construct Synchronized Network Graphs to aid in our analysis and provide a means for visual examination. Finally, we performed analysis of inorganic vs. organic synchronization within the networks by overlaying bot detection analysis. These three steps are described in detail in this section.

\subsection*{Calculation of Combined Synchronization Index for each Event}
From the raw Twitter data for each of the six events, we first find synchronous users by identifying users who perform the same action within a five-minute time window. We use the synchronous user data to calculate CSI-UserPair values for each unique user pair within the network. We next use the CSI-UserPair data to calculate CSI-User for each unique user in the dataset, before averaging the CSI-User values to obtain CSI-Network. 

We use three action types in the event calculation: hashtags, URLs and @mentions. Using multiple action types provides a view of different groups that synchronize across different dimensions, resulting in a more holistic quantification of the combined synchronization that is happening within the network. 

We also evaluate the CSI-Network using only one action type, i.e. CSI-Network formulated from hashtag action type, CSI-Network formulated from URL action type etc. Comparing CSI-Network formulated from singular action types against multiple action types reveals the advantage of combining networks. The sequence of steps taken for the construction of a single action type is the same as for multiple action types, except that we evaluate the CSI-UserPair, CSI-User and CSI-Network separately for each action type, resulting in three CSI-Network values, one for each action type.

\subsection*{Construction of Synchronized Network Graphs}
From the CSI-UserPair data, we generate synchronized network models. We construct a weighted, undirected user-user synchronization network $G = (V,E)$. Each node $v\in V$ corresponds to a synchronizing user. An edge $E_{ij}$ between users $v_i$ and $v_j$ represents the degree of sychronity of the two users $i$ and $j$. The weight of the edge, $w_{e_{ij}}$ is the value of the CSI-UserPair. From the graph, we can calculate the network density, which is the ratio of the number of formed edges to the number of possible edges for the network. Visually, users are represented as circular nodes, edges as lines between nodes and the thickness of the lines represent the edge weights. The visual density and clusters of the network graphs provide supplemental analysis towards the quantitative scores.

\subsection*{Inorganic vs. Organic Synchronization}
We define inorganic synchronization as that stemming from inauthentic bot users; and organic from human users. 
Inorganic synchronization can serve to amplify certain themes in the discourse. Bots can also team up with each other to change and influence the viewpoints of human users.
Organic synchronization can be worrying because human users that agree with each other on harmful trends such as \#stopthesteal can possibly spillover into real-world protests when the narrative is echoed frequent enough.

We use the BotHunter algorithm to differentiate the authenticity of a user. The algorithm uses a random forest model with posts and user metadata features for user classification. It returns a bot likelihood score, which we deem users with bot likelihoods above 0.70 as a bots and humans otherwise \cite{ng2022stabilizing}.

With this differentiation of users into bot/human user types, we calculated the average CSI-User by user type, which indicates which users are likely to synchronize in which event. We also break down the average CSI-UserPair by user pair type -- bot-bot, bot-human and human-human -- type. This breakdown can tell us which pair of user group perform more synchronization, allowing possible inferences towards the trajectory of the discourse of the event.

Finally, we overlay these bot classification on our Synchronized Network Graphs constructed in the previous step to aid in visual analysis of the authenticity of synchronicity within each event.

\subsection*{Network Metrics Analysis}
Users coordinate with each other within a social network, hence we also investigate the social structures of interaction using network centrality metrics derived from graph theory knowledge. These network centrality measures characterize the users' positions within a social network graph, and by extension, their activity and behavior with other users.

Within this work, we measure three network centrality values: total degree centrality, eigenvector centrality and betweenness centrality. Total degree centrality represents the total number of edges a user node has. The higher the degree, the more the number of connections the user has, and the more central the user is. The eigenvector centrality value calculates the influence of a node not only based on its connections but also based on the centrality values of its connected neighbors. The higher the eigenvector centrality value of a user, the more the user is connected to nodes that are highly central, and the higher the user's influence in the network. Betweenness centrality is based on the graph-theoretic measure of shortest paths. This centrality measure calculates the extent to which a user lies on the shortest path that connects two other users. The higher the betweenness centrality of a user, the more the user has influence over the flow of information within a network graph. We measure these three centrality values of the users at each step of the CSI hierarchy.

The first step of the CSI hierarchy formulates CSI-UserPair. For analyzing the centrality metrics of users by their participation in user pairs coordination. We examine the sets of user pairs that coordinate across 1, 2 and 3 action types, we use a user-to-user all-communication network graph. 

A user-to-user all-communication network graph, $A=(B,D)$, is formulated in the following manner. Each node $b\in B$ corresponds to a user that we have collected a tweet during our data collection. An edge $D_{ij}$ between users $b_i$ and $b_j$ is formed when the two users interacted with each other, i.e. retweet, quote, mentions. The weight of the edge $w_{D_{ij}}$ represents the number of interactions between the two users $b_i$ and $b_j$. For simplicity, we formulate this as an un-directed graph because the number of links across the graph can be quite extensive.

We use this network graph formation in the calculation of user centrality as it captures all the raw interactions between user-pairs. The extent and the temporal drift of these interactions are what we extract to measure user synchronity in our CSI measures.

In the next step, as we calculate the CSI-User values, we also correlate them to user centrality values in the network graphs. Here, we also use the user-to-user all-communication graphs for measurement of centrality values. We compare the centrality values of bot and human users that participate in synchronous activities, examining the difference in synchronity and influence within the network between the two user types.

Finally, we evaluate the modularity and hierarchy values of the resultant Synchronized Network Graphs, which we had described their construction in the previous step. Modularity is the measure of the structure of the graph and its ability to be sub-divided into smaller clusters of users. The lower the modularity, the more interconnected the network is, which means more synchronicity is occurring between the users within the network graph. The hierarchy values of a network describe the ability of a network to be divided into smaller social structures. The small structure a user node can be embedded in is a dyad, which are two user nodes joined together. The lower the hierarchy values, the harder it is to extract increasingly larger groups of social groups of users from the network, which means that the network contains more synchronicity. Both measures are of the range of 0 to 1.

We also compare the bot/human partitions of the Synchronized Network Graphs in terms of their global clustering coefficient, or transitivity \cite{watts1998collective}. This measure ranges from 0 to 1. This measures the tendency of the nodes in the network to cluster together. The higher the global clustering score, the more the network contains groups of nodes that are loosely connected to each other and will form clusters on a network visualization. We compare these scores to the dominant partition that visible on the resultant network graphs, providing both quantitative and qualitative measures of synchronized network action.

\section*{Results and Discussion}
We present the results of the evaluation of synchronization with our hierarchical Combined Synchronized Index (CSI) across the six events. These results are presented in terms of the results and analysis of synchronization across user pairs, users and networks. We also further evaluated centrality features of each step of the hierarchy, providing insights to trends between user network centrality values and their likelihood to synchronize.

\subsection*{Analyzing User Pair Synchronization.}
User pairs were examined for their synchronization across three action types: URLs, hashtags and @mentions. Across all events, an average of approximately 89\% of the users coordinate across only one action type, (Table \ref{tab:numuserspercoordination}). This lends weight to the synchronization metric used that it captures more intentional coordination rather than coincidental synchronization of users. We infer that users that synchronize across multiple action types are likely to be performing deliberate coordination in which they harmonize their posts and the post artifacts within.


\begin{table*}[!hpt]
\centering
\begin{tabular}{|p{3.2cm}|p{2.5cm}|p{3cm}|p{2.5cm}|}
\hline
& \multicolumn{3}{|c|}{\textbf{\# Action Types users coordinate across}} \\ \hline
\textbf{Event} & \textbf{1 Action Type} & \textbf{2 Action Type} & \textbf{3 Action Type} \\ \hline 
Black Panther 2018 & \textbf{0.61} & 0.31 & 0.08\\ \hline 
CharlieHebdo shooting 2020 & \textbf{0.93} & 0.07 & 0.005\\ \hline 
Reopen America 2020 & \textbf{0.96} & 0.03 & 0.01 \\ \hline 
COVID Vaccine 2021 & \textbf{0.94} & 0.05 & 0.01 \\ \hline 
US Elections Primaries 2020 & \textbf{0.96} & 0.02 & 0.02\\ \hline
Capitol Riots 2021 & \textbf{0.96} & 0.03 & 0.01\\ \hline \hline
Average & \textbf{0.89} & 0.09 & 0.02 \\ \hline
\end{tabular}
\caption{Percentage of users against the number of action types they coordinate}
\label{tab:numuserspercoordination}
\end{table*}

We perform a deeper examination of the differences between the users that synchronize across the number of action types. This examination is done with network centrality measures, which characterize the users' positions in the social network graph. In performing this calculation, the original user-to-user all-communication graph was used, which provides information about the users' positions with respect to all the interactions they performed during the event.

\begin{figure}[h!]
\includegraphics[width=1.0\textwidth]{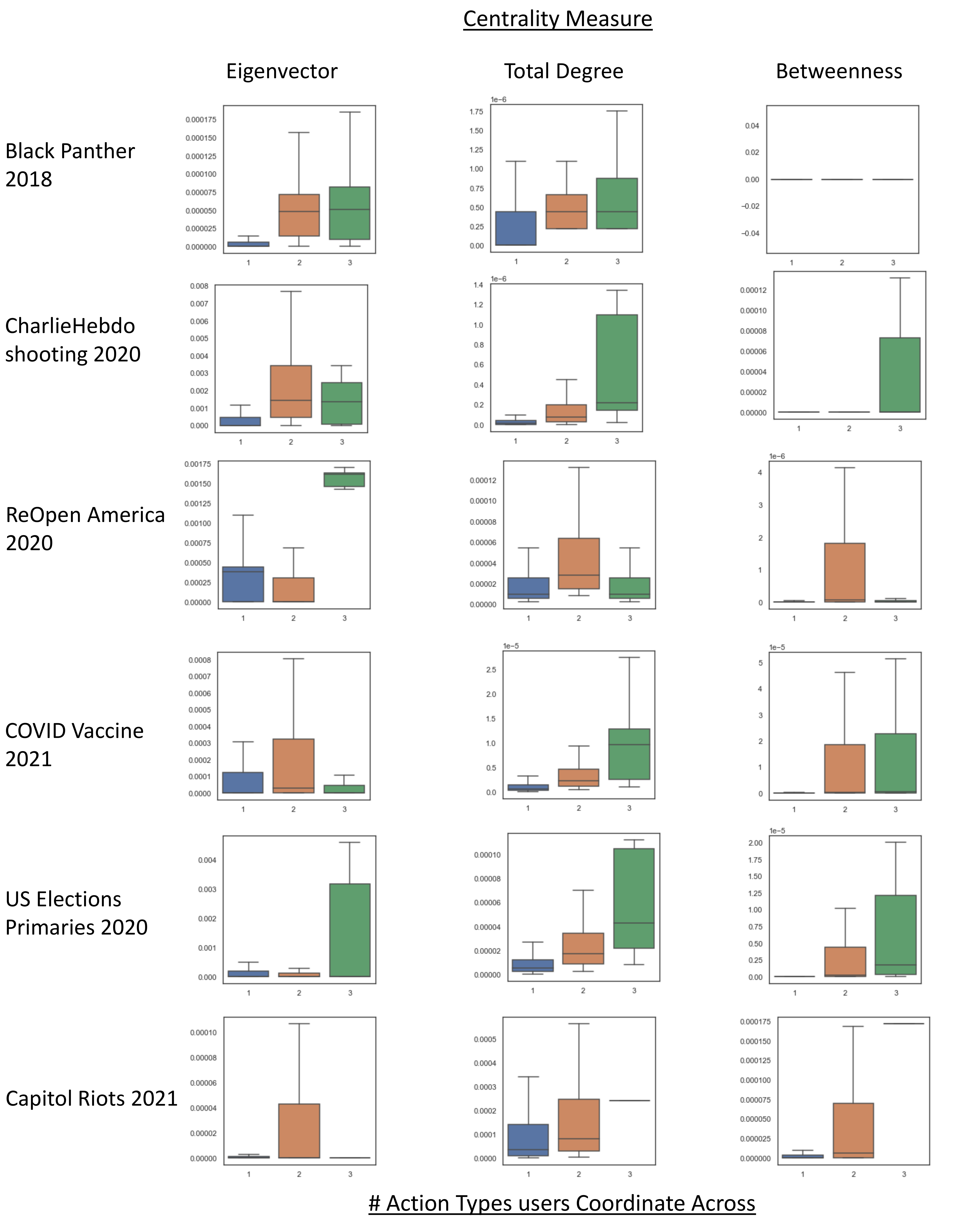}
\caption{\csentence{Analyzing User Pairs with Network Centrality Measures} Boxplot representations of network centrality measures against the number of action types users coordinate across. These graphs represent user-to-user all-communication graphs.}
\label{fig:user_centrality}
\end{figure}

The results of the number of action types a user perform against its centrality measures are plotted in \autoref{fig:user_centrality}. In particular, we are interested in users that synchronized across all three action types. We observe that in terms of centrality measures, these users generally have the highest mean with respect to the measure. This means that users that are highly active in synchronizing across multiple actions are typically more central: higher total degree centrality means they have more connections; higher eigenvector centrality means they have more influence; higher betweenness centrality means they lie on the shortest paths of other users. Intuitively, this provides face validity: users that are synchronizing with other users are more likely to have more connections and interactions. With more connections, these synchronizing users can receive a larger spread of information, and can also disseminate their information further and wider among the network. We also observe that users that synchronize across three types also have large standard deviations, which reflects that users can coordinate across several action types irregardless of their position in the interaction network. 

Using information about whether a user is bot or human, we characterized the user pairs into Bot-Bot, Human-Human and Bot-Human synchronization pairs. The full statistics of the average CSI-UserPair for each network is displayed in Table \ref{tab:userpaircoord}. 

Across all events, we consistently observe that the average CSI-UserPair is highest in the bot-human pair, in which one user is a bot and one user is a human. This indicates that the bots are successful within the network, given that they are designed to engage with humans \cite{bessi2016social}. The observation of synchronization of bots and humans indicates that the two classes are working together to push a common narrative, revealing coordinated activity within the network. Coupled with previous work which shows a high engagement of humans with bots \cite{schuchard2019bots}, the higher synchronization values between bot-human pairs show that the two classes of users can coordinate with each other across physical boundaries bridged by the social media platforms.

A high CSI-UserPair value between bot-human users indicate either play of inauthentic activity influencing human users, or of bots mimicking human users in the online space. Bot users can influence the online conversation through extensive coordination techniques during the event, resulting in higher bot-human CSI-UserPair scores. This can be worrying because excessive exposure of humans to malicious bot activity can cause them to change their opinion, especially towards discourses unfavorable towards the social public. This synchronization exchange between bot-human pairs should be carefully watched, because only 5-10\% of bots echoing the same topic in the network is sufficient to tilt the balance of the public opinion towards the bots' desired opinion \cite{CHENG2020124163,ross2019social}. This opens avenues for further investigation into the influence pattern between the two user types.

Bots can also mimic and potentially manipulate humans in the online space \cite{ferrara2016rise,alothali2018detecting}. Early bots and simplistic bots operate by mimicking human behavior, though recent observations have documented high coordination among groups of bots to direct attention to specific information \cite{ferrara2016rise}. This characteristic have been observed in political and election campaigns, collectively disseminating specific political hashtags in increased volumes during the campaign season  
\cite{keller2019social}.

Bots acting like humans in online conversations can potentially cause a series of harms, ranging from psychological to social harms \cite{daniel2019bots}. Some instances that have been observed are bots making death threats \cite{singleton_2015} and bots negatively affecting or overwhelming political discussions \cite{bessi2016social}. 

Therefore, the presence of inorganic synchronization and synchronization between inauthentic users is an important part to investigate, due to the possibilities of the actions affecting human psychology and society.

\begin{table*}[!hpt]
\centering
\begin{tabular}{|p{3.2cm}|p{2.5cm}|p{3cm}|p{2.5cm}|}
\hline
\textbf{Event} & \textbf{Bot-Bot} & \textbf{Human-Human} & \textbf{Bot-Human} \\ \hline 
Black Panther 2018 & 1.17 & 1.20 & \textbf{1.23}\\ \hline 
CharlieHebdo 2020 & 1.04 & 1.20 & \textbf{1.23}\\ \hline 
ReOpen America 2020 & 1.04 & 0.99 & \textbf{1.05} \\ \hline 
COVID Vaccine 2021 & 1.01 & 1.01 & \textbf{1.02} \\ \hline 
US Elections Primaries 2020 & 1.08 & 1.10 & \textbf{1.10} \\ \hline
Capitol Riots 2021 & 1.07 & 1.15 & \textbf{1.18} \\ \hline \hline
Average & 1.07 & 1.11 & \textbf{1.15} \\ \hline
\end{tabular}
\caption{Average CSI-UserPair by user-user pair types}
\label{tab:userpaircoord}
\end{table*}

\subsection*{Analyzing User Synchronization}
From CSI-UserPair, we derive CSI-User through a weighted aggregation of user synchronicity. Table \ref{tab:usercoordination} shows the statistics of CSI-User, differentiated by user type within each of the six events. On average, the amount of synchronization between human users are more than that of bot users (11.92 vs 10.19). However, closer examination of individual events reveal differences between the dominant synchronizing user class. Most of the events reveal that humans participate in the synchronization between each other more than bot users except for ReOpen America and COVID Vaccine release events. We infer that bots do not typically target smaller scale social activism events like the Black Panther and CharlieHebdo events, but target large-scale events such as ReOpen America event. Past studies show that humans are primarily responsible for the content generation. In our study, we are only examining original tweets which are responsible for content generation, leading to our observations that humans synchronize with each other more during these events. Humans thus are coordinating to generate content, and it is likely that bots use their automated mechanisms to amplify the content through retweeting behavior \cite{kitzie2018life}. 

Bots were not actively synchronizing with each other in original posts during the CharlieHebdo shooting event. This mirrors studies on shootings where posts contributed by bots can outpace humans in some events but are lag behind in other events \cite{schuchard2019bots}. It is thus an open research topic to identify the events that bots take an interest in.

In the Capitol Riots event, humans reveal an average CSI-User that is almost twice that of the bots, which indicates some users being extremely passionate about the event, and were likely coordinating the movement on the US Capitol Hill. Similarly, the US Election Primaries has a high synchronicity despite it being a huge and broad topic. This indicates that there are some users that are passionate about American politics and were likely coordinating in their topical discussions in hopes to swing the electoral vote.

\begin{table*}[!hpt]
\centering
\begin{tabular}{|p{4cm}|p{3cm}|p{3cm}|}
\hline
\textbf{Event} & \textbf{Bots} & \textbf{Human} \\ \hline
Black Panther 2018 & 2.14$\pm$0.97 & \textbf{2.77$\pm$1.75} \\ \hline 
CharlieHebdo shooting 2020 & 4.12$\pm$2.84 & \textbf{4.18$\pm$5.18} \\ \hline 
ReOpen America 2020 & \textbf{7.37$\pm$5.39} & 7.22$\pm$5.42 \\ \hline
COVID Vaccine release 2021 & \textbf{7.62$\pm$22.76} & 5
90$\pm$6.06\\ \hline
US Elections Primaries 2020 & 32.45$\pm$22.55 & \textbf{38.16$\pm$29.26} \\ \hline
Capitol Riots 2021 & 7.44$\pm$7.24 & \textbf{13.29$\pm$10.41} \\ \hline \hline
Average & 10.19$\pm$11.13 & \textbf{11.92$\pm$13.36} \\ \hline 
\end{tabular}
\caption{Average CSI-User by User Type}
\label{tab:usercoordination}
\end{table*}

Using the initial all-communication networks, we calculated the average centrality values of bots/human classes. These values are presented in \autoref{tab:usercentrality}, separated per event. Together with the CSI-User scores, centrality value tells us about the patterns of organic and inorganic synchronization occurring within the events.

Overall, we observe that bots are more centrally placed than human users. The bot accounts have a higher total degree centrality than humans, signifying that they form more communication connections than human users. They also have higher eigenvector centrality values compared to human users, signifying their higher influence in the network and their connectivity to other high influence users. Lastly, the higher betweenness centrality values show that these bots lie on the paths of information dissemination and act as information brokers or transmissions.

The higher centrality values of bots compared to humans lends weight to the fact that many bot accounts are designed to manipulate online conversations and hence place themselves central to the conversations through excessive and possibly calculated interactions. This is in line with past studies where as little as 4\% of well-placed bot accounts is sufficient to overturn the majority opinion \cite{ross2019social}. Together with the observations of the average CSI-User by bot/human user type, we infer that bots place themselves at central positions, and thus are able to actively synchronize their actions in large scale events such as the ReOpen America and COVID Vaccine events. In these events, their synchronization can easily manipulate the network because they are already well-positioned to do so. These two events also culminated in offline consequences: ReOpen America protests and anti-vaccination sentiments. While we are unable to entirely attribute the offline activity to online synchronization, a part of the human stance must have been influenced by the synchronous online activity.

\begin{table*}[!hpt]
\centering
\begin{tabular}{|p{3.5cm}|p{1.5cm}|p{1.5cm}|p{1.5cm}|p{1.5cm}|p{1.5cm}|p{1.5cm}|}
\hline
\textbf{Event} & \multicolumn{3}{|c|}{\textbf{Bots}} & \multicolumn{3}{|c|}{\textbf{Humans}}  \\ \hline
& Total Degree Centrality & Betweenness Centrality & Eigenvector Centrality & Total Degree Centrality & Betweenness Centrality & Eigenvector Centrality \\ \hline
Black Panther 2018 [Humans] & 1.28E-6 & 6.00E-6 & 3.65E-4 & 1.67E-6 & 2.00E-6 & 1.78E-4 \\ \hline
CharlieHebdo shooting 2020 [Humans] & 2.61E-7 & 9.07E-4 & 2.83E-3 & 6.87E-8 & 1.56E-4 & 9.56E-4 \\ \hline
ReOpen America 2020 [Bots] & 2.90E-5 & 9.08E-4 & 2.83E-3 & 2.00E-5 & 4.61E-6 & 1.79E-4 \\ \hline
COVID Vaccine release 2021 [Bots] & 2.03E-6 & 4.70E-6 & 1.46E-3 & 1.72E-6 & 5.04E-6 & 5.14E-4 \\ \hline 
US Elections Primaries 2020 [Humans] & 1.97E-5 & 3.98E-6 & 4.15E-3 & 4.73E-5 & 2.50E-6 & 1.76E-3 \\ \hline
Capitol Riots 2021 [Humans] & 1.42E-4 & 1.53E-5 & 2.53E-3 & 5.40E-5 & 1.10E-5 & 5.52E-4 \\ \hline 
\hline
Average & 3.24E-5 & 3.07E-4 & 2.36E-3 & 2.08E-5 & 3.02E-5 & 6.90E-4 \\ 
\hline
\end{tabular}
\caption{Network centrality measures by bot/human classes that participate in synchronous activities. We opted not to present the standard deviations as the values are extremely small ($<10^{-6}$). The user class type in [] beside the event name indicates the user class with the higher average CSI-User.}
\label{tab:usercentrality}
\end{table*}

\subsection*{Analyzing Network Synchronization}
\begin{figure}[h!]
\includegraphics[width=1.0\textwidth]{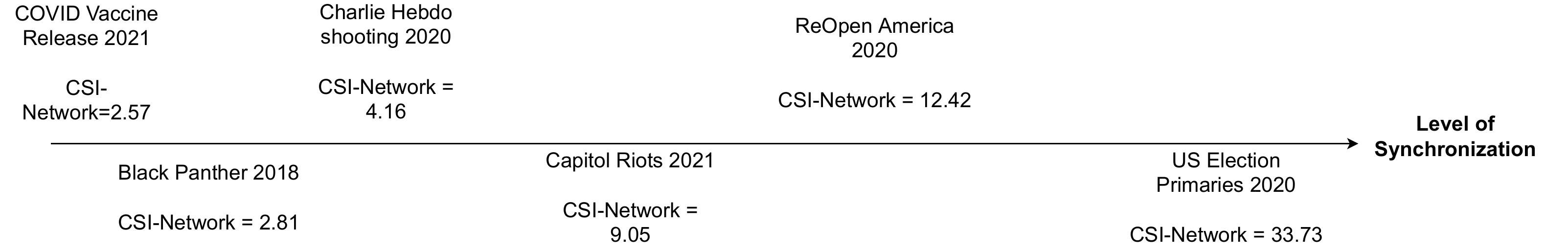}
\caption{\csentence{Level of Synchronization}
  \revision{Events ordered by their level of synchronization, from least to most.}}
\label{fig:csi_line}
\end{figure}

The level of user synchronization and coordination within the six events provides CSI-Network scores for the events that ranges from 2.57 to 33.73 (Table \ref{tab:csinetwork_indiv_full}). The CSI-Network formulation is able to separate the events along a scale to represent user coordination. The events, ordered by the user coordination levels are: COVID Vaccine Release 2021, Black Panther 2018, CharlieHebdo shooting 2020, Capitol Riots 2021, ReOpen America 2020 and US Elections Primaries 2020. Figure \ref{fig:csi_line} places the six events on a line scale in the order of their level of synchronization.

\begin{figure*}
    \centering
    \subfigure[\textbf{Black Panther 2018} \newline CSI-Network=2.81 \newline Network density=1.61E-3 ]{\includegraphics[width=0.45\textwidth]{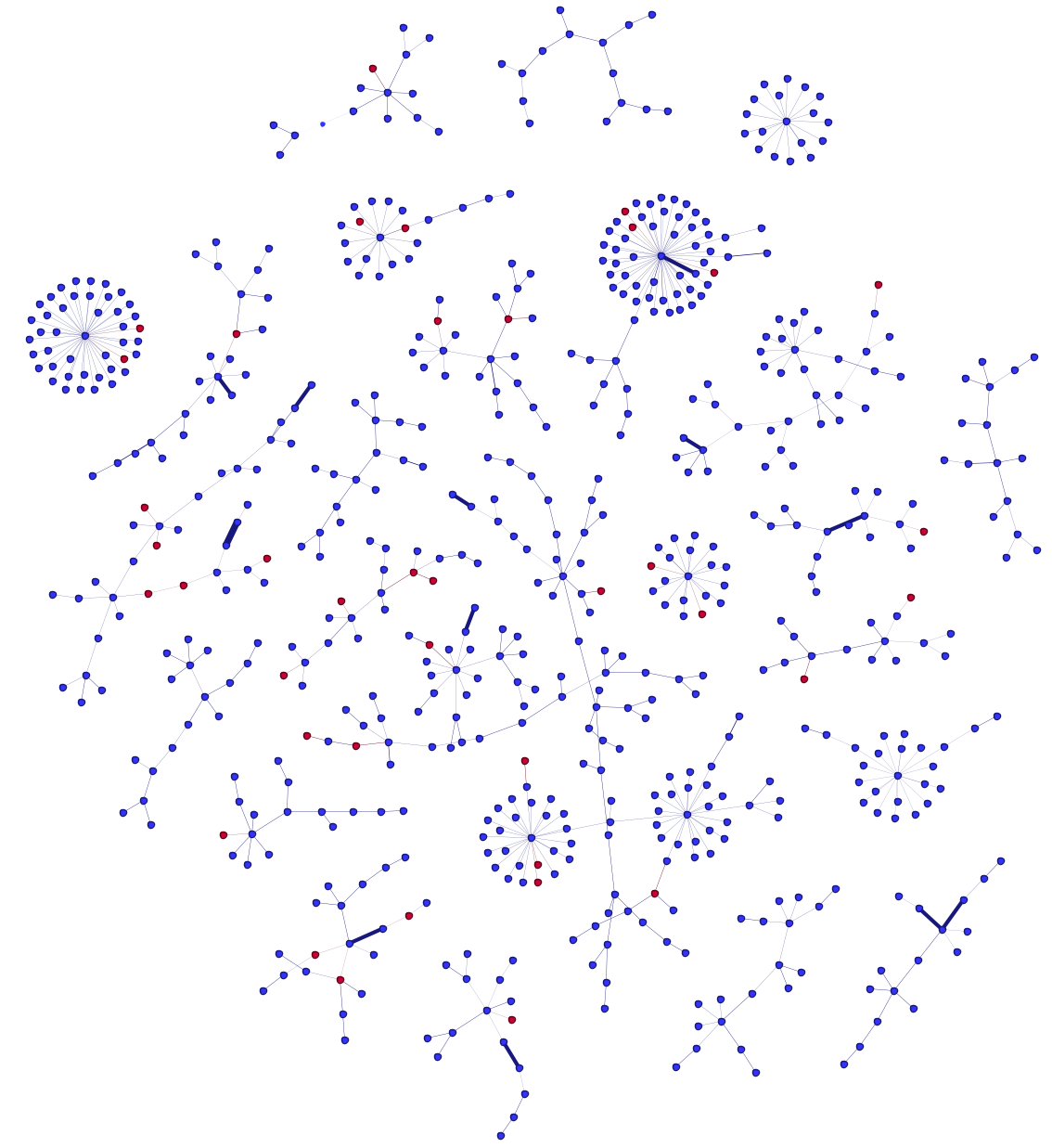}}
    \subfigure[\textbf{CharlieHebdo shooting 2020} \newline CSI-Network=4.16 \newline Network density=4.8E-5]{\includegraphics[width=0.45\textwidth]{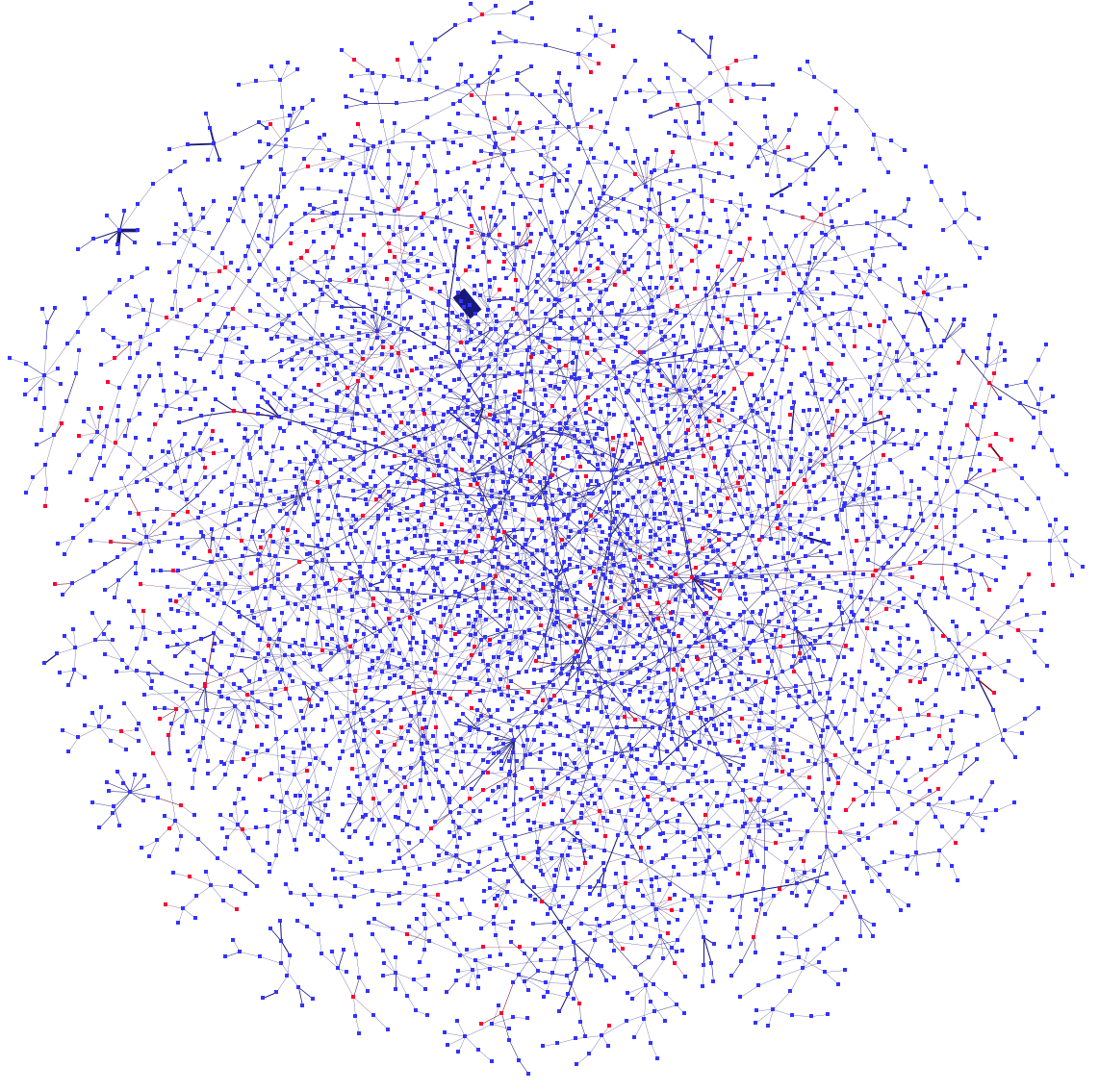}}
    \subfigure[\textbf{ReOpen America 2020} \newline CSI-Network=12.42 \newline Network density=1.59E-4]{\includegraphics[width=0.45\textwidth]{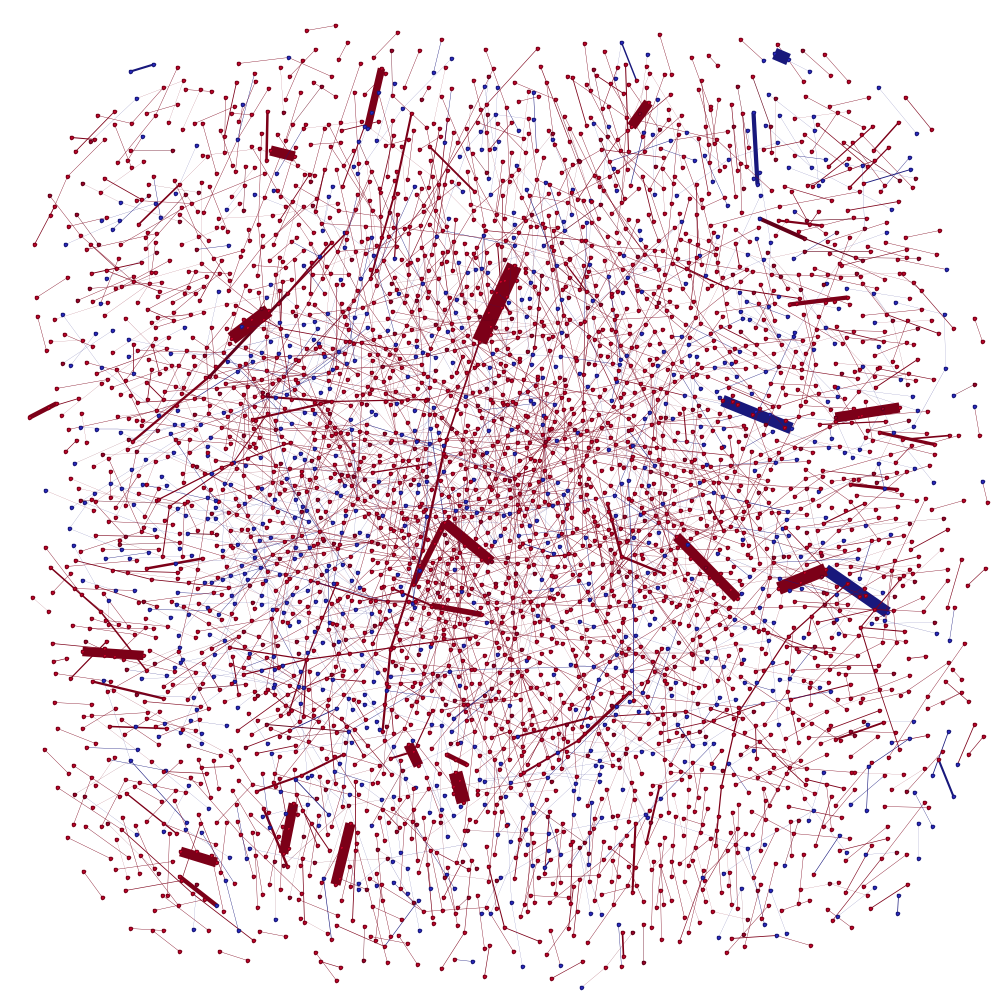}}    
    \subfigure[\textbf{COVID Vaccine Release 2021} \newline  CSI-Network=2.57 \newline Network density=1.69E-5]{\includegraphics[width=0.45\textwidth]{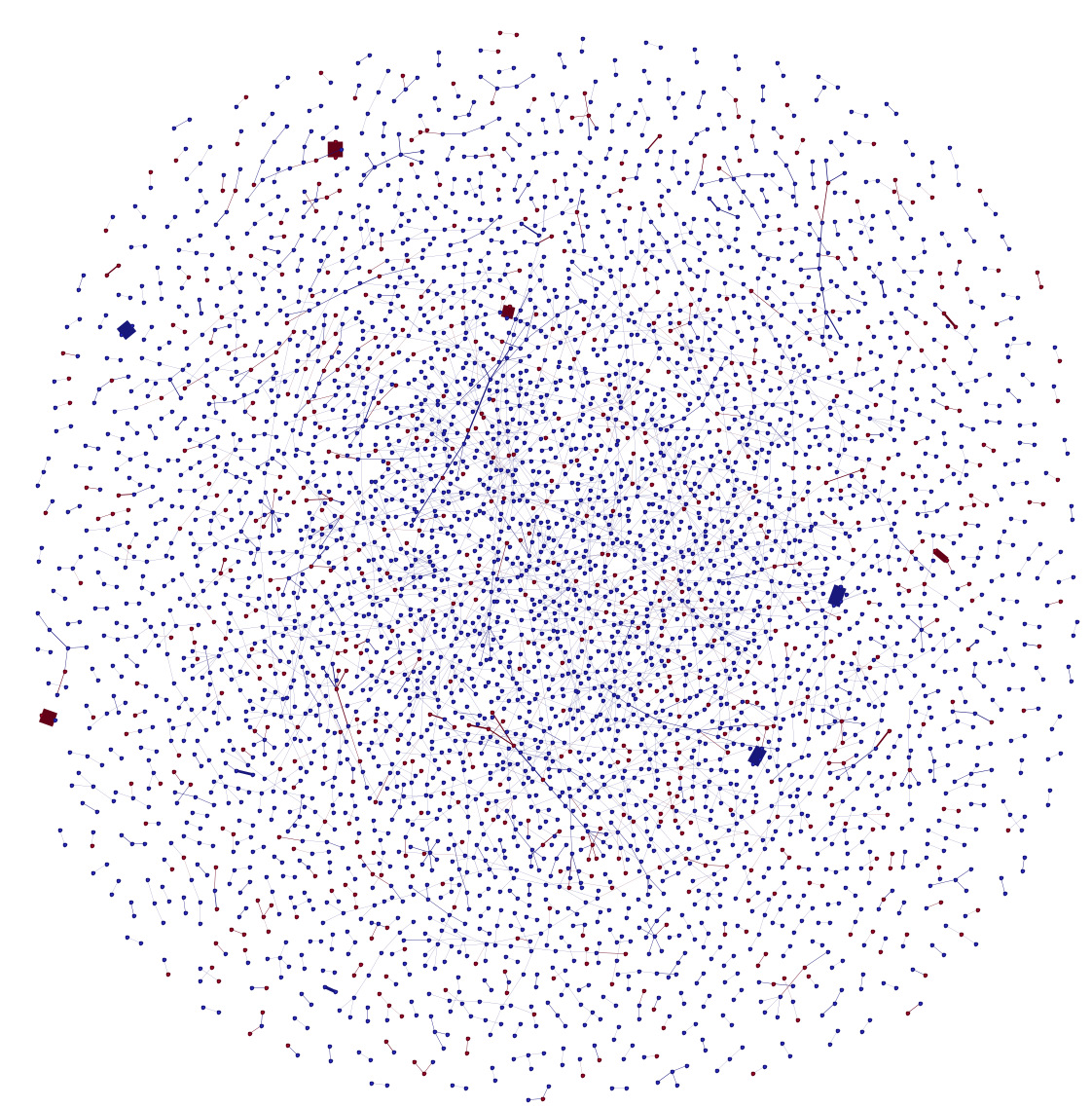}} 
    \subfigure[\textbf{US Elections Primaries 2020} \newline  CSI-Network=33.73 \newline Network density = 8.90E-4]{\includegraphics[width=0.45\textwidth]{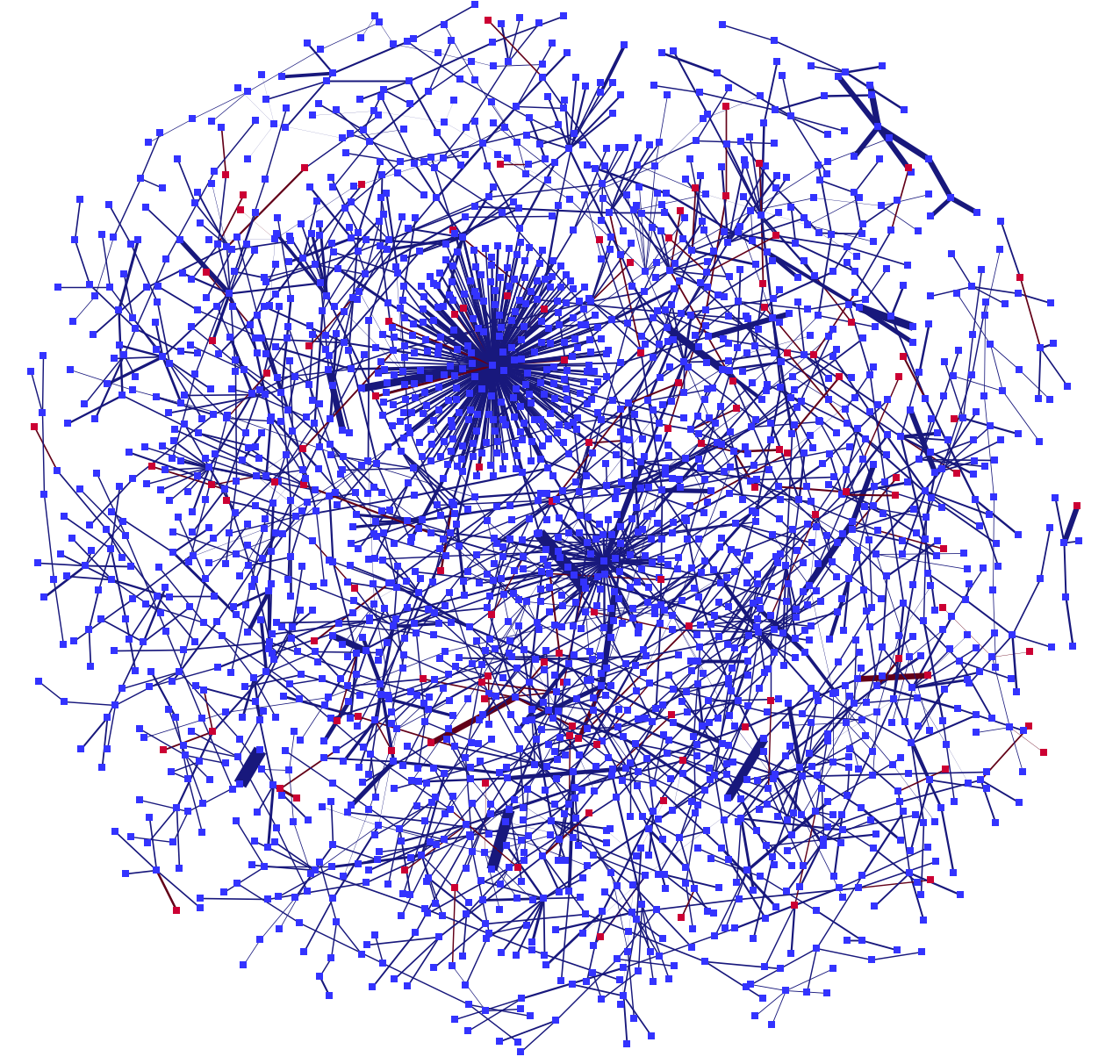}}
    \subfigure[\textbf{Capitol Riots 2021} \newline CSI-Network=9.05 \newline Network density=9.7E-4]{\includegraphics[width=0.45\textwidth]{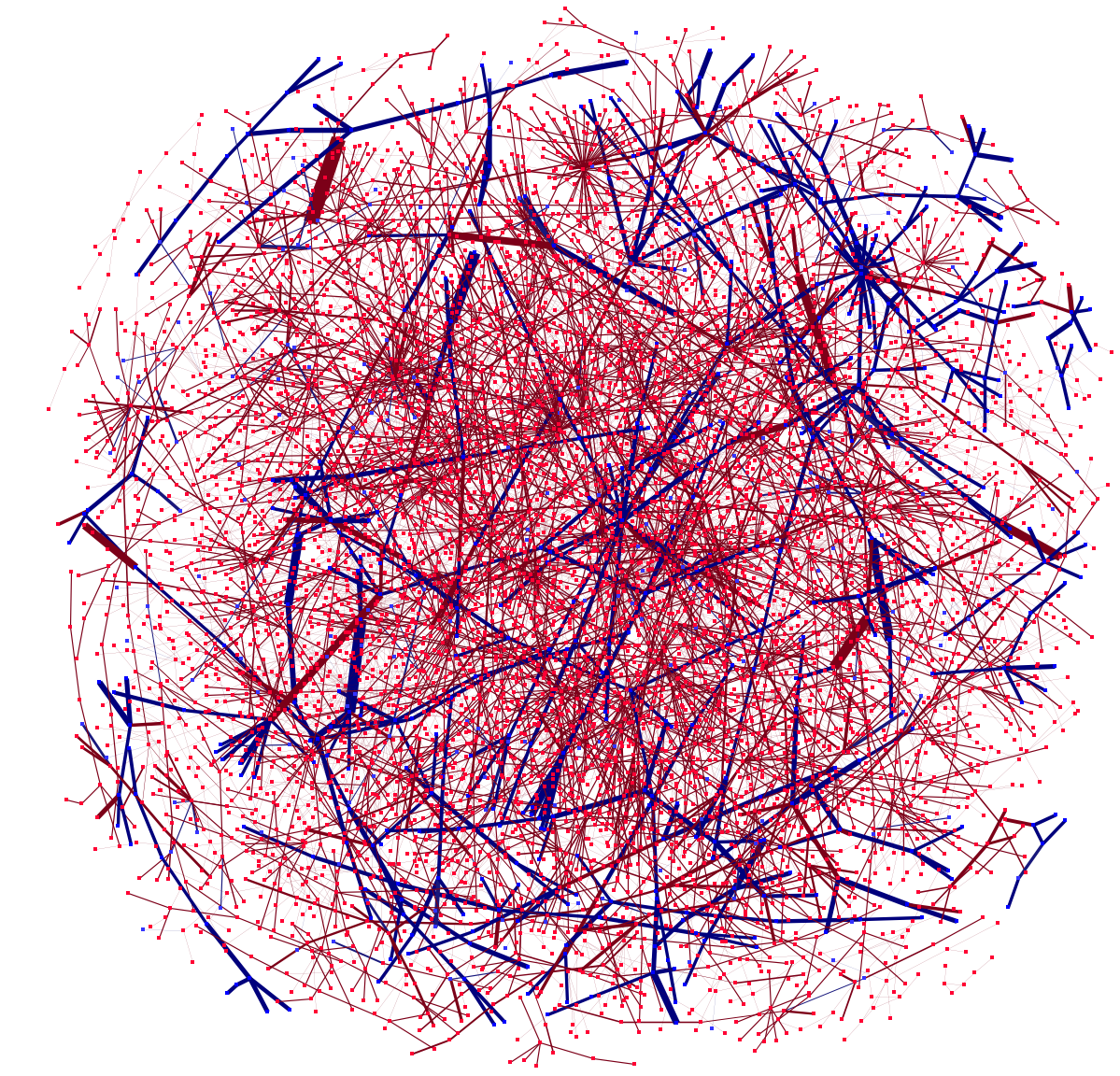}} 
    \caption{\csentence{Synchronized Network Graphs.} Nodes are users. Red nodes are bots and blue nodes are humans. Links two users represent synchronization between them. Link widths represent the degree of synchronization. Graphs have been pruned to show nodes that synchronize with at least 5 different users to depict only the core structure of users that synchronize very frequently.}
    \label{fig:networkgraphs}
\end{figure*}

Visualizing the synchronization network further suggests that higher CSI-Network scores correlate to denser networks, measured by the network density. The order of the networks in accordance to the network density lines up with the CSI-Network scores, revealing that the CSI-Network quantitatively supports the network graph as a measure of synchronicity. Figure \ref{fig:networkgraphs} shows the Synchronized Network Graphs constructed from the CSI-UserPair data. 


We note that the topic of US Elections and ReOpen America are higher impact and more longstanding conversational topics, and are therefore likely to have a large number of people expressing their opinion on it online, resulting in higher levels of coordination, a part of it could be coincidental synchronicity.  

Lastly, the dominant color of the network graphs indicates the bot participation within the events, thereby indicating the extent the event is organic. Graphs that are predominantly red exhibit bots participate in synchronization more than humans, while graphs that are predominantly blue exhibit humans participate in synchronization more. The two network graphs that are predominantly red are the ReOpen America 2020 and Capitol Riots 2021 graphs. Besides the intense online discussion, these two events resulted in offline protests and riots. The presence of synchronized inauthentic behavior within these two events could possibly be a factor to the spillover of the real-world as riots and protests respectively. The bots could be amplifying and spreading the narratives supporting and inciting violence in a coordinated fashion. Users with high CSI-UserPair scores, depicted by the thicker lines in the network graphs, can be extracted and monitored over time. Their activity traces will give analysts glimpses of changes in narratives that can better help prepare for any social movements. 

The other events show higher coordination between human users, and the extent of coordination reveals the interest humans pay in each event. A huge political event such as the US Elections garners more interests than a social activism event like Black Panther or a shooting. Within social activism events, events that have clear sides like the COVID Vaccine Release with its pro- and anti-vaccine stances seem to have more conversation as compared to the shooting event. 

We note that some datasets may have conversations predominantly in another language and are thus not captured by our English language filter and downstream analysis. For example, the Charlie Hebdo shooting dataset might have dominanted the French Twittersphere during the incident, but are not captured in our dataset. We expound on this in the discussion on the limitations of this study.

We further note that there are differences between the conclusions in the CSI-User and CSI-Network analysis for the COVID vaccine release and Capitol Riots events. For the COVID vaccine event, the CSI-User scores reveals that bots synchronize more than humans, while its synchronized network visualization shows a higher human likeliness within the network graph. The Capitol Riots event reveals the converse case. The CSI-User scores looks at the synchronization within the individual level, while the network visualization provides a larger network perspective. Closer examination of the networks of the COVID vaccine release reveals clusters of bots with high CSI-User scores which can contribute to the observation of higher synchronization across bots. The standard deviation of the CSI-User for bots is extremely high at 22.76 points in the COVID vaccine event, which indicates there is a wide spectrum of CSI-User scores that likely contributed to skewing the average score. The same obervations can be made of the Capitol Riots event. 

\subsection*{Network Centrality and Synchronization}
The six events examined are political and social activism events. Their CSI-Network revealed a spectrum of degree of synchronization within the events, showing the extent of synchronization and coordination within these events. This section further examines the final synchronized networks formed and the centrality scores of the network as a whole, as well as partitions of the network (bot, human groups).

\autoref{tab:csinetwork_modularity} shows the comparison of CSI-Network scores against network modularity metrics. These network modularity metrics are derived from the resultant synchronized network graphs. In these graphs, nodes represent users that participate in synchronous activities, and links between two nodes represent that the two users had performed a synchronous action. Overall, the synchronous network graphs formed have high modularity and hierarchy values, irregardless of the density of the graphs. The users that are synchronizing with each other are very tightly coupled together, which is consistent with the design of the definition of synchronization.

The global clustering coefficient scores for the resultant networks, though, tell a different story. For many events, a low CSI-Network score corresponds to a low global clustering coefficient score, which reflects that the resultant networks does not form clusters very well. Indeed, if we were to examine the Black Panther event, it has both a low CSI-Network and global clustering coefficient score, which corresponds a rather disjointed network with few clusters, which is visualized in \autoref{fig:networkgraphs}. However, events like the COVID Vaccine release, US Election Primaries and Capitol Riots have CSI-Network and global clustering coefficient scores that are on two opposite ends of the spectrum. This inconsistency of scores points us to investigate into the network further, as it indicates that there are some partitions of the network that are more clustered than others.

\begin{table}[!hpt]
\centering
\begin{tabular}{|p{1.8cm}|p{1.8cm}|p{1.8cm}|p{1.8cm}|p{1.8cm}|p{1.8cm}|}
\hline
\textbf{Event} & \textbf{CSI-Network} & \textbf{Newmann Modularity} & \textbf{Krackhardt's Hierarchy} & \textbf{Density} & \textbf{Global Clustering Coefficient} \\ \hline
Black Panther 2018 & 2.81 & 0.998 & 1 & 1.61E-3 & 0.119 \\ \hline
CharlieHebdo shooting 2020 & 4.16 & 0.999 & 0.999 & 4.8E-5 & 0.512 \\ \hline 
ReOpen America 2020 & 12.42 & 0.996 & 1 & 1.59E-4 & 0.535 \\ \hline 
COVID Vaccine release 2021 & 2.57 & 0.994 & 1 & 1.69E-4 & 0.580 \\ \hline 
US Elections Primaries 2020 & 33.73 & 0.995 & 1 & 8.90E-4 & 0.223 \\ \hline 
Capitol Riots 2021 & 9.05 & 0.992 & 0.999 & 9.70E-4 & 0.783 \\ \hline 
\end{tabular}
\caption{Comparison of CSI-Network scores against Network Modularity metrics derived from the Synchronized Network Graphs}
\label{tab:csinetwork_modularity}
\end{table}

We then investigate the resultant synchronized networks in terms of the partition of bot/human classes. \autoref{tab:csinetwork_transitivity} describes the global clustering coefficient of each group. Here, we observe consistency between the global clustering coefficient scores and the dominant bot/human partition group that from qualitative analysis of network visualization. This reflects the group that has high tendency to form clusters in the resultant network, which indicates that they have higher synchronicity among each other. This information, coupled with the qualitative visual analysis, provides insight to whether the synchronization was organic or inorganic.

For example, the global clustering coefficient of the network formed for Capitol Riots 2021 is 0.783, which is extremely high. In contrast, its CSI-Network is 9.05, which is on the lower end of the spectrum. However, when we examine the clustering coefficients of bot and human partitions separately, we see that that bot partition has a clustering coefficient of 0.785, while the human users have a clustering coefficient of 0.245. Therefore, during the Capitol Riots event, the bots were actively synchronizing with each other, resulting in high clustering formations in the resultant synchronized network graphs. In contrast, human users synchronize lesser in all multiple dimensions, resulting in the lower CSI-Network scores formulation.

Through using a combination of CSI-Network score, overall global clustering coefficients and clustering coefficients of network partitions, we can deduce whether online synchronization that happens during an event is largely organic or inorganic. This harmony, and lack of, between CSI-network scores and network centrality metrics can be harnessed to observe the presence of organic/inorganic synchronization. This observation provides us information where bots or humans are more engaged in synchronization during an event. Hopefully, this information can potentially aid analysts in preparing for the possibility of offline violence.

\begin{table}[!hpt]
\centering
\begin{tabular}{|p{3.5cm}|p{1.8cm}|p{1.5cm}|p{1.5cm}|p{1.8cm}|}
\hline
\textbf{Event} & \textbf{CSI-Network} & \textbf{Bot} & \textbf{Human} & \textbf{Dominant Group in Visualization} \\ \hline
Black Panther 2018 & 2.81 & 0.569 & \textbf{0.998} & Human \\ \hline
CharlieHebdo shooting 2020 & 4.16 & 0.177 & \textbf{0.381} & Human \\ \hline 
ReOpen America 2020 & 12.42 & \textbf{0.621} & 0.267 & Bot \\ \hline 
COVID Vaccine release 2021 & 2.57 & 0.330 & \textbf{0.634} & Human \\ \hline 
US Elections Primaries 2020 & 33.73 & 0.216 & \textbf{0.234} & Human \\ \hline 
Capitol Riots 2021 & 9.05 & \textbf{0.785} & 0.245 & Bot \\ \hline 
\end{tabular}
\caption{Comparison of CSI-Network Scores against Global Clustering Coefficient scores derived from the Synchronized Network Graphs. The scores are split up by bot/human classes and are consistent with the dominant group in the network graph visualizations.}
\label{tab:csinetwork_transitivity}
\end{table}

\subsection*{Using Multiple Action Types for Synchronization Analysis}
In all the above analyses, we looked at the construction of CSI-Network values using multiple action types -- hashtags, mentions and URLs. Using multiple action types balances out any distortions in the synchronization index due to skews in the action types taken by users. Table \ref{tab:csinetwork_indiv_full} shows the comparison between CSI-Network values constructed out of a single action type vs the values constructed with multiple action types. We obtain one CSI-Network value per action type and compare that against the value constructed from multiple action types.

We generally observe that CSI-Network using Mentions, CSI-Network (Mentions), are the lowest, suggesting that users typically do not @mention other users within an original tweet. We observe a higher value in two cases -- Black Panther 2018 and CharlieHebdo shooting 2020 -- possibly indicating users are tagging other users to bring attention to the event.

CSI-Network constructed using Hashtags, CSI-Network (Hashtag), have the highest value among the three action types. Users typically suffix their tweet texts with hashtags, sometimes using multiple hashtags, thus allowing a wider surface area to be detected by the synchronization calculation. Events that are larger in scope like ReOpen America garner larger attention and have more commonly used hashtags and thus have higher synchronization values. The high values of synchronization among CSI-Network for URLs, CSI-Network (URLs), indicates that URLs are a commonly used artifact to coordinate information sharing among users. 

These individual CSI-Network values, however, are not representative of the full nature of synchronization across the event. If one were to only examine the CSI-Network (Hashtag) value of 23.29 for the Black Panther 2018 event, one might conclude that the user synchronization within that event is way more than that of the CharlieHebdo shooting 2020 event (CSI-Network (Hashtag) = 2.06). However, if one compared the CSI-Network (Mentions) of both events, the conclusion would be the opposite, i.e. CharlieHebdo shooting is more coordinated than BlackPanther event. Thus, multiple action types must be considered to smoothen out excessive synchronization in one dimension by single users. In a combination of action types, users that participate in synchronization across multiple action types are given higher weights. The CSI-Network thus reflects the proportion of users that synchronize in multiple dimensions, possibly deliberately coordinating in the message spread. This aids in identifying users that are truly coordinating: synchronizing in high frequencies across multiple dimensions.

\begin{table}[!hpt]
\centering
\begin{tabular}{|p{4cm}|p{2.0cm}|p{2.0cm}|p{1.8cm}|p{1.8cm}|}
\hline
\textbf{Event} & \textbf{CSI-Network} & \textbf{CSI-Network} & \textbf{CSI-Network} & \textbf{CSI-Network} \\ \hline 
\textbf{Action Type} & \textbf{Hashtag} & \textbf{Mentions} & \textbf{URLs} & \textbf{All Three} \\ \hline
Black Panther 2018 & 23.29 & 2.07 & 10.49 & 2.81\\ \hline
CharlieHebdo shooting 2020 & 2.06 & 3.02 & 1.02 & 4.16\\ \hline 
ReOpen America 2020 & 49.10 & 1.03 & 5.89 & 12.42\\ \hline 
COVID Vaccine release 2021 & 7.06 & 1.09 & 19.62 & 2.57 \\ \hline 
US Elections Primaries 2020 & 37.69 & 1.01 & 39.30 & 33.73\\ \hline 
Capitol Riots 2021 & 7.76 & 1.03 & 38.28 & 9.05\\ \hline 
\end{tabular}
\caption{Comparison of CSI-Network Scores for one action type vs multiple action types}
\label{tab:csinetwork_indiv_full}
\end{table}

%


\subsection*{Limitations and Future Work.}
Several limitations nuance our conclusions from this work. In terms of data collection, the Twitter API retrieval technique used returns only a 1\% sample of Tweets. Our datasets are curated by a sampling of the Twitter platform via selected keywords and timeframe, and so only show a subset of the full interactions. There may be synchronizing users that are not captured in the datasets. Further work on understanding synchronization across languages other than English is also required, in order to better map out the synchronization behavior of users across different languages. Additionally, our operationalization of synchronization only considers actions in isolation, i.e. creating posts with the same hashtag or the same URL. It could be expanded to investigate hierarchies of actions, i.e. creating posts with \textit{both} the same hashtag and URL within the specified time period. 

Future work involves expanding the CSI to factor in user synchronization across moving time windows. This provides a way to perform effective comparisons of user synchronization and the nature of synchronization across different time periods. The nature of our dataset collection are across short time spans of weeks, and further collection work will be required to expand the dataset into months for longitudinal time series analysis. This will provide a view of the evolution of the online discourse across time. Further, this provides insight on whether inauthentic bot users are artificially driving the change, or whether it is an organic movement that should be closely monitored. Another direction of research is to look at synchronization of post artifacts, i.e. hashtags, URLs, across platforms. While there are preliminary studies on cross-platform URL synchronization, the studies primarily examined which URLs are most shared among a users of the multiple platforms, rather than quantifying whether the users are coordinating their sharing of URLs within and across platforms \cite{ng2022cross,phadke2020many}.

\section*{Conclusions}
Synchronization of users within the online social media environment is a key factor affecting online conversations. The cacophony of users performing actions simultaneously reveals some form of deliberate coordination, which can provide clues to the subsequent trends of the conversation, and can preempt any offline protests or riots. 

The present work introduces the Combined Synchronization Index (CSI) to quantify the degree of synchronization within social network discourse. This tool serves as a qualitative, network-centric method that abstracts the inter-dimensional perspective of synchronization across key social media post artifacts: @-mentions, hashtags and URLs. There are no significant differences between the CSI scores of all three hierarchies of political and social activism events, hence the CSI formulation can be said to be generalizable within this scope. The CSI is a three-tiered hierarchy index which characterizes the synchronization between user pairs (CSI-UserPair), the participation of a user in concurrence with other users (CSI-User), and the overall user synchronization within a network of conversation derived from an event (CSI-Event). 

In this work, six political and social activism events were analyzed and characterized using the Combined Synchronization Index. These event types are chosen as they are of huge impact and have the potential to turn into offline protests and riots. The CSI separates the events in accordance to their level of synchronicity, and further analyzes the user interactions within the event. We observe that in a top-down view of the events, there is a dominant user type (i.e. bot or human) that participate in synchronous activities. Human users have a higher CSI-User score, indicating that they have a higher participation of synchronization with original posts than bots. The trend of bot-human pairs having the highest synchronization index across all events is concerning as it reveals that bots do have an effect on human users, and can possibly manipulate their opinions.

Limitations aside, socio-technical systems can benefit from the quantification of synchronization between users within events. Understanding the degree of synchronization and which user class types participate in synchronizing actions within each event can aid in predicting the intensity of the discussion surrounding the event. This index serves as a general way to quantify synchronization across different dimensions: posting with the same hashtag, posting with the same URL, or even posting with the same hashtag and URL; allowing for comparison of overall event synchronization statistics instead of singular ones. We hope this work can serve as a jumping point for further investigation into synchronized and coordinated efforts on social media platforms.  



\begin{backmatter}

\section*{Abbreviations}
CSI: Combined Synchronization Index

URL: Uniform Resource Locator

\section*{Competing interests}
The authors declare that they have no competing interests.

\section*{Author's contributions}
LHXN: Conceptualization, methodology, analysis, writing. KMC: Conceptualization, review and editing. All authors have read and agreed to the published version of the manuscript. 

\section*{Acknowledgements}
The research for this paper was supported by the following grants: Cognizant Center of Excellence Content Moderation Research Program, Office of Naval Research (Bothunter, N000141812108), Scalable Technologies for Social Cybersecurity/ARMY (W911NF20D0002), Air Force Research Laboratory/CyberFit (FA86502126244). The views and conclusions  are those of the authors and should not be interpreted as representing the official policies, either expressed or implied.  

\section*{Availability of data and materials}
The datasets that were generated and analyzed in this study are available upon request from the corresponding author. The code can be found at \url{https://github.com/quarbby/synchronization_index}.


\bibliographystyle{bmc-mathphys} 
\bibliography{bmc_article}      

\end{backmatter}
\end{document}